\documentclass{eptcs}
\usepackage{tikz}
\usepackage{amsmath,amsfonts,amssymb,amsthm}
\usepackage{listings}
\usepackage{verbatim}
\usepackage{alltt}

\theoremstyle{definition}
\newtheorem{exmp}{Example}

\usepackage{breakurl}             % Not needed if you use pdflatex only.

\def\ddef{\mathop{\tt ::=}}

\def\ti{\mbox{\tt i}}
\def\ts{\mbox{\tt s}}
\def\tc{\mbox{\tt c}}
\def\tf{\mbox{\tt f}}

\def\eps{\mbox{\tt eps}}

\def\prholog{P$\rho$Log}

\newtheorem{example}{Example}

\title{P$\rho$Log: a system for rule-based programming}
\author{Besik Dundua
	\institute{Kutaisi International University\\ Kutaisi, Georgia}
	\institute{Ilia Vekua Institute of Applied Mathematics, 
	 Ivane Javakhishvili Tbilisi State University\\
		Tbilisi, Georgia}
	\email{bdundua@gmail.com}
}

\begin{document}
	\maketitle
	
%	\begin{abstract}
%		This is a sentence in the abstract.
%		This is another sentence in the abstract.
%		This is yet another sentence in the abstract.
%		This is the final sentence in the abstract.
%	\end{abstract}
	
	\section{Brief overview}
	
	P$\rho$Log \cite{ DBLP:conf/padl/DunduaKR17} is a rule-based system that supports programming with individual, sequence, function and context variables. It extends Prolog with rule-based programming capabilities to manipulate sequences of terms. The four kinds of variables help to traverse tree forms of expressions both in horizontal and vertical directions, in one or more steps. It facilitates to have expressive pattern matching that often helps to write short and intuitive code.
	
	Another important feature of P$\rho$Log is the use of strategies. They provide a mechanism to control complex rule-based computations in a highly declarative way. With the help of strategies, the user can combine simpler transformation rules into more complex ones. In this way, P$\rho$Log conveniently combines the whole Prolog power with rule-based strategic programming features.  
	
	P$\rho$Log is based on $\rho$Log calculus \cite{DBLP:journals/jancl/MarinK06}, where the inference system basically is the SLDNF-resolution with normal logic program semantics~\cite{DBLP:books/sp/Lloyd87}. It has been successfully used in the extraction of frequent patterns from data mining workflows \cite{nguyen2015meta}, XML transformation and web reasoning \cite{DBLP:conf/rr/CoelhoDFK10}, modeling of rewriting strategies \cite{DBLP:journals/corr/abs-1001-4434} and access control policies \cite{DBLP:conf/sacmat/MarinKD19}, etc.  
	
	The $\rho$Log calculus has been influenced by the $\rho$-calculus~\cite{DBLP:journals/igpl/CirsteaK01,DBLP:journals/igpl/CirsteaK01a}, which, in itself, is a foundation for the rule-based programming system ELAN~\cite{DBLP:journals/entcs/BorovanskyKKMV96}. There are some other languages for programming by rules, such as, e.g.,  ASF-SDF~\cite{DBLP:journals/entcs/BrandDHJJKKMOSVVV01}, CHR~\cite{DBLP:journals/jlp/Fruhwirth98}, %Claire~\cite{DBLP:journals/tplp/CaseauJL02}, 
	Maude~\cite{DBLP:journals/tcs/ClavelDELMMQ02}, Stratego~\cite{DBLP:conf/rta/Visser01}, Tom~\cite{DBLP:conf/rta/BallandBKMR07}. The $\rho$Log calculus and, consequently, {\prholog} differs from them, first of all, by its pattern matching capabilities. Besides, it adopts logic programming semantics (clauses are first class concepts, rules/strategies are expressed as clauses) and makes a heavy use of strategies to control transformations. Earlier works about $\rho$Log and its implementation in Mathematica include~\cite{MarinK2003,DBLP:journals/entcs/MarinP04,DBLP:conf/synasc/MarinI05}.
	
	P$\rho$Log is available at \url{https://www.risc.jku.at/people/tkutsia/software/prholog}.
		
	\section{The P$\rho$Log language}
	
	We write P$\rho$Log constructs in \texttt{typewriter} font. They are \emph{terms} (including a special kind of terms, called \emph{contexts}) and sequences. These objects are constructed from function symbols without a fixed arity (so called unranked or variadic function symbols), a special constant $\tt hole$ (called the \emph{hole}), and individual, functional, context and sequence variables. These variables are denoted by the identifiers whose names start respectively with $\verb#i_# ,\verb#f_#,\verb#c_#$, and $\verb#s_#$ (e.g., $\verb#i_X# ,\verb#f_X#,\verb#c_X#, \verb#s_X#$).
Terms \texttt{t} and sequences \texttt{s} are constructed in a
standard way:
$$\begin{array}{ll}
	\tt t \ \ddef \ {\tt hole} \mid {\tt \ti\_X}\mid f(s)\mid {\tt \tf\_X(s)\mid \tc\_X}(t) &\text{(terms)}\\
	\tt s \ \ddef \ {\tt \eps}\mid t\mid {\tt \ts\_X}\mid (s,s) &\text{(sequences)}
\end{array}$$
where $\tt f$ is a function symbol and $\tt eps$ stands for the empty sequence and is omitted whenever it appears as a subsequence of another sequence. A \emph{context} is a term with a single occurrence of the $\tt hole$ constant. Application of a context $\tt C$ to a term $\tt t$ is a term derived by replacing the hole in $\tt C$ with $\tt t$. For instance, applying $\verb#f(i_X,g(i_Y,hole),a)#$ to $\verb#g(b,hole)#$ gives another context $\verb#f(i_X,g(i_Y,g(b,hole)),a)#$, while applying it to  $\verb#g(b,c)#$ gives a non-context term $\verb#f(i_X,g(i_Y,g(b,c)),a)#$.

A \emph{substitution} is a mapping from individual variables to hole-free terms, from sequence variables to hole-free sequences, from function variables to function variables and symbols, and from context variables to contexts, such that all but finitely many individual, sequence, and function variables are mapped to themselves, and all but finitely many context variables are mapped to themselves applied to the  \verb#hole#. This mapping can be extended to terms and sequences in the standard way. For instance, for a substitution
 $\sigma=$\verb#{c_Ctx#$\mapsto$\verb#f(hole),# \verb#i_Term#$\mapsto$\verb#g(s_X)#, \verb#f_Funct#$\mapsto$\verb#g#, \verb#s_Seq1#$\mapsto$\verb#eps,# \verb#s_Seq2#$\mapsto$\verb#(b,c)}# and a sequence \verb#s#$=$\verb#(c_Ctx(i_Term),f_Funct(s_Seq1,a,s_Seq2))#, by applying $\sigma$ to $\verb#s#$ we get the sequence $\sigma($\verb#s#$)=$ \verb#(f(g(s_X)),g(a,b,c)).#

\emph{Matching problems} are pairs of sequences, one of which is ground (i.e., does not contain variables). Such matching problems may have zero, one, or more (finitely many) solutions, called matching substitutions or \emph{matchers}. For instance, the sequence \verb#(s_1,f(i_X),s_2)# matches \verb#(f(a),f(b),c)# in two different ways: one by the matcher \verb#{s_1#$\mapsto$\verb#(),i_X#$\mapsto$\verb#a,s_2#$\mapsto$\verb#(f(b),c)}# and other one by the matcher \verb#{s_1#$\mapsto$\verb#f(a),i_X#$\mapsto$\verb#b,# \verb#s_2#$\mapsto$\verb#c}.# Similarly, the term \verb#c_X(f_Y(a))# matches the term \verb#f(a,g(a))# with the matchers \verb#{c_X#$\mapsto$\verb#f(hole,g(a)),f_Y#$\mapsto$\verb#f}# and \verb#{c_X#$\mapsto$\verb#f(a,g(hole)),f_Y#$\mapsto$\verb#g}.#  Matching is the main computational mechanism in {\prholog}.

Instantiations of sequence and context variables can be restricted by regular sequence and regular context languages, respectively. We do not go into the details of this feature of P$\rho$Log matching here.
%These constraints are expressed as \verb#s_X in# $\rh$ and \verb#c_X in# $\rc$, where  $\rh$ and  $\rc$ are regular hedge and context expressions defined by the grammars:
%{\small
%	\begin{alignat*}{1}
%		\rh ::={} & \mathtt{eps} \mid (\rh\ \rh) \mid \rh|\rh \mid \rh^* \mid \texttt{f}(\rh) \mid \rc(\texttt{f}(\rh)) \\
%		\rc ::={} & \mathtt{hole} \mid \rc.\rc \mid \rc+\rc \mid \rc^\star \mid \texttt{f}(\rh,\rc,\rh)
%\end{alignat*}}%
%For $\rh$, juxtaposition stands for concatenation, the vertical bar $\vert$ for choice, and $^*$ for repetition. For $\rc$, the dot is concatenation, $+$ is choice, and $^\star$ is repetition.	
	
A $\rho$Log\ \emph{atom} ($\rho$-atom) is a quadruple consisting of a hole-free term $\tt st$ (a \emph{strategy}), two hole-free sequence $\tt s1$ and $\tt s2$, and a set of regular constraints {\tt R} where each variable is constrained only once, written as \verb#st :: s1 ==> s2 where R#. Intuitively, it means that the strategy $\tt st$ transforms $\tt s1$ to $\tt s2$ when the variables satisfy the constraint {\tt R}. We call $\tt s1$ the left hand side and $\tt s2$ the right hand side of this atom. When {\tt R} is empty, we omit it and write \verb#st :: s1 ==> s2#. The negated atom is written as \verb#st :: s1 =\=> s2 where R#. A $\rho$Log \emph{literal} ($\rho$-literal) is a $\rho$-atom or its negation. A P$\rho$Log \emph{clause} is either a Prolog clause, or a clause of the form  \verb#st :: s1 ==> s2 where R :- body# (in the sequel called a $\rho$-clause) where \texttt{body} is a (possibly empty) conjunction of $\rho$- and Prolog literals.

A P$\rho$Log{} \emph{program} is a sequence of P$\rho$Log\ clauses and a \emph{query} is a conjunction of $\rho$- and Prolog literals. There is a restriction on variable occurrence imposed on clauses: $\rho$-clauses and queries can contain only $\rho$Log variables, and Prolog clauses and queries can contain only Prolog variables. If a Prolog literal occurs in a $\rho$-clause or query, it may contain only $\rho$Log individual variables that internally get translated into Prolog variables.

\section{Inference and strategies}

{\prholog} execution principle is based on depth-first inference with leftmost literal selection in the goal. If the selected literal is a Prolog literal, then it is evaluated in the standard way. If it is a {\prholog} atom of the form $\verb#st :: s1 ==> s2#$, due to the syntactic restriction called well-modedness (formally defined in \cite{DBLP:journals/corr/abs-1001-4434}), $\verb#st#$ and $\verb#s1#$ do not contain variables. Then a (renamed copy of a) program clause $\verb#st' :: s1' ==> s2' :- body#$ is selected, such that the strategy $\verb#st'#$ matches $\verb#st#$ and the sequence \verb#s1'# matches $\verb#s1#$ with a substitution $\sigma$. Next, the selected literal in the query is replaced with the conjunction $\sigma(\verb#body#), \verb#id :: #\sigma(\verb#s2'#)\verb# ==> s2#$, where $\verb#id#$ is the built-in strategy for identity: it succeeds iff its right-hand side matches the left-hand side. Evaluation continues further with this new query. Success and failure are defined in the standard way. Backtracking explores other alternatives that may come from matching the selected query literal to the head of the same program clause in a different way (since context/sequence matching is finitary, see, e.g.,~\cite{DBLP:journals/jsc/Comon98a,DBLP:conf/lpar/KutsiaM05}), or to the head of another program clause. Negative literals are processed by negation-as-failure.

When instead of the exact equality one uses proximity as in~\cite{dundua2019extending}, then in place of $\verb#id#$, {\prholog} introduces another built-in strategy in the new query: $\verb#prox#(\lambda)$, which succeeds if its right-hand side matches the left-hand side approximately, at least with the degree $\lambda$.\footnote{This is an experimental feature, not yet included in the official distribution.} Proximity relations indicate by which degree two expressions are close to each other, where the degree is a real number in $[0,1]$. Proximity with degree 1 means that the terms are equal, while degree 0 means that they are distinct. Hence, $\verb#id#$ can be seen as an abbreviation of $\verb#prox#(1)$. Proximity is a fuzzy reflexive, symmetric, non-transitive relation, suitable for modeling imprecise, incomplete information.

Some of the other predefined strategies of {\prholog} and their intuitive meanings are the following:
\begin{itemize}
%	\item \verb#id :: h1 ==> h2# succeeds if the hedges \verb#h1# and \verb#h2# are identical (or can be made identical by \verb#h2# matching \verb#h1#) and fails otherwise.
	\item $\mathtt{compose(st_1,st_2,\ldots,st_n)}$, $n\ge 2$, first transforms the input sequence by $\mathtt{st_1}$ and then transforms the result by $\mathtt{compose(st_2,\ldots,st_n)}$ (or by $\mathtt{st_2}$, if $n=2$). Via backtracking, all possible results can be obtained. The strategy fails if either $\mathtt{st_1}$ or $\mathtt{compose(st_2,\ldots,st_n)}$ fails.

	\item $\mathtt{choice(st_1,\ldots,st_n)}$, $n\ge 1$, returns a result of a successful application of some strategy $\mathtt{st_i}$ to the input sequence. It fails if all $\mathtt{st_i}$'s fail. By backtracking it can return all outputs of the applications of each of the strategies $\mathtt{st_1,\ldots,st_n}$.

	\item \verb#first_one#$\mathtt{(st_1,\ldots,st_n)}$, $n\ge 1$, selects the first $\mathtt{st_i}$ that does not fail on the input sequence and returns only one result of its application. \verb#first_one# fails if all $\mathtt{st_i}$'s fail. Its variation, \verb#first_all#, returns via backtracking all the results of the application to the input sequence of the first strategy $\mathtt{st_i}$ that does not fail.
%	\item $\mathtt{nf(st)}$, when terminates, computes a normal form of the input hedge with respect to \texttt{st}. It never fails because if an application of \texttt{st} to a hedge fails, then \texttt{nf(st)} returns that hedge itself. Backtracking returns all normal forms.

%	\item $\mathtt{iterate(st, \mathit{N})}$ starts transforming the input hedge with \texttt{st} and returns a result (via backtracking all the results) obtained after $N$ iterations for a given natural number $N$.

	\item $\mathtt{map(st)}$ maps the strategy \texttt{st} to each term in the input sequence and returns the result sequence. Backtracking generates all possible output sequences. \texttt{st} should operate on a single term and not on an arbitrary sequence. $\mathtt{map(st)}$ fails if \texttt{st} fails for at least one term from the input sequence. 

%	\item \texttt{interactive} takes a strategy from the user, transforms the input hedge by it and waits for further user instruction (either to apply another strategy to the result hedge or to finish).
%	\item $\mathtt{rewrite(st)}$ applies to a single term (not to an arbitrary hedge) and rewrites it by \texttt{st} (which also applies to a single term). Via backtracking, it is possible to obtain all the rewrites. The input term is traversed in the leftmost-outermost manner. Note that $\mathtt{rewrite(st)}$ can be easily implemented inside \prholog{}: \\
%	\verb#   rewrite(i_Str) :: c_Context(i_Redex) ==> c_Context(i_Contractum) :-#\\
%	\verb#        i_Str :: i_Redex ==> i_Contractum.#
\end{itemize}

\section{Examples}

In this section we bring some examples to illustrate features and the expressive power of {\prholog}.

\begin{exmp}[Sorting]
	The following program illustrates how bubble sort can be implemented in P$\rho$Log.
\begin{verbatim}
   swap(f_Ordering) :: (s_X, i_I, i_J, s_Y) ==> (s_X, i_J, i_I, s_Y) :- 
       not(f_Ordering(i_I, i_J)).
   bubble_sort(f_Ordering) := first_one(nf(swap(f_Ordering))).
\end{verbatim}

In the first clause, the user-defined strategy $\verb#swap#$ swaps two neighboring elements $\verb#i_I, i_J#$ in the given sequence if they violate the given ordering $\verb#f_Ordering#$. The use of sequence variables $\verb#s_I, s_J#$ helps to identify the violating place in the given sequence by pattern matching, without the need to explicitly define the corresponding recursive procedure. The sorting strategy $\verb#bubble_sort#$ is then defined as an exhaustive application of $\verb#swap#$ (via the built-in strategy $\verb#nf#$), which will lead to a sorted sequence. The strategy $\verb#first_one#$ guarantees that only the first answer computed by $\verb#nf(swap(f_Ordering))#$ is returned: it does not make sense to sort a sequence in different ways to get the same answer over and over again via backtracking.

The way how the $\verb#bubble_sort#$ strategy is defined above is just an abbreviation of the clause
\begin{verbatim}
   bubble_sort(f_Ordering) :: s_X ==> s_Y :- 
       first_one(nf(swap(f_Ordering))) :: s_X ==> s_Y.
\end{verbatim}

{\prholog} allows the user to write such abbreviations. To use this strategy for sorting a sequence, we could call, e.g.,
\begin{verbatim}
   ?(bubble_sort(=<) :: (1,3,4,3,2) ==> s_X, Result).
\end{verbatim}
and {\prholog} would return a single result in the form of a substitution:
\begin{verbatim}
   Result = [s_X ---> (1,2,3,3,4)].
\end{verbatim}
\end{exmp}\vspace{0.15cm}

%\begin{comment}

\begin{exmp}[Rewriting] One step of rewriting a term by some rule/strategy can be straightforwardly defined in {\prholog}:
\begin{verbatim}
    rewrite_step(i_Str) :: c_Ctx(i_X) ==> c_Ctx(i_Y) :- i_Str :: i_X ==> i_Y.
\end{verbatim}

It finds a subterm $\verb#i_X#$ in $\verb#c_Ct(i_X)#$ to which the strategy $\verb#i_Str#$ applies and rewrites it. Due to the built-in matching algorithm that finds the relevant instantiation of the context variable $\verb#c_Ctx#$, this step corresponds to a leftmost-outermost rewriting step. In \cite{DBLP:journals/corr/abs-1001-4434} we have illustrated how easily other rewriting strategies can be modeled in {\prholog}.
\end{exmp}\vspace{0.15cm}

\begin{comment}
\begin{example}\label{rewwith}{\em One-step rewriting with regular constraint.
		\begin{verbatim}
			rewrite_under_f(i_Str) :: c_X(i_X) ==>
			    c_X(i_Y) where ([c_X in c_(f(h_,hole,h_))]) :-
		            	i_Str :: i_X ==> i_Y.
		\end{verbatim}
		
		Now, we can see how P$\rho$Log rewrites terms using rewriting
		defined by Example~\ref{rewwithout} and Example~\ref{rewwith}
		with respect to one-rule rewriting system $a \rightarrow b$.
		\begin{verbatim}
			?(rewrite_one_step(st):: f(f(g(a),a),a) ==> h_x,Result).
			Result = [h_x ---> f(f(g(b), a), a)] ;
			Result = [h_x ---> f(f(g(a), b), a)] ;
			Result = [h_x ---> f(f(g(a), a), b)] ;
			false.
			
			?(rewrite_under_f(st):: f(f(g(a),a),a) ==> h_x,Result).
			Result = [h_x ---> f(f(g(a), a), b)] ;
			Result = [h_x ---> f(f(g(a), b), a)] ;
			false.
		\end{verbatim}

}\end{example}

\end{comment}

\begin{exmp}[Using proximity] We assume that a proximity relation between function symbols is given. (Based on it, we can compute proximity between terms as well.) The task is to remove from a given sequence approximate duplicates, i.e., if the sequence contains two elements that are proximal to each other (by a predefined degree), we should get a sequence where only one of the proximal elements is retained. The strategy $\verb#merge_proximals#$ below does it. It checks whether the sequence contains somewhere two elements $\verb#i_X#$ and $\verb#i_Y#$ that are close to each other at least (by the given degree) and removes $\verb#i_X#$. $\verb#merge_all_proximals#$ removes all such approximate duplicates and returns one answer:
\begin{verbatim}
   merge_proximals :: (s_X, i_X, s_Y, i_Y, s_Z) ==> (s_X, s_Y, i_Y, s_Z) :-
       prox :: i_X ==> i_Y.
   merge_all_proximals := first_one(nf(merge_proximals)).
\end{verbatim}

Assume our proximity relation is such that $\verb#a#$ and $\verb#b#$ are proximal with degree 0.6 and $\verb#b#$ is close to $\verb#c#$ with degree 0.8. (Every symbol is proximal to itself with degree 1.) Then we can ask to merge all proximals in the sequence \verb#(a,b,d,b,c)# first by degree 0.5 and then by degree 0.7. The resulted answers show the result sequence and the actual approximation degree (which is $\ge$ the given degree):
\begin{verbatim}
   ?(merge_all_proximals :: (a,b,d,b,c) ==> s_Ans, 0.5, Degree, Result).
   Degree = 0.6,
   Result = [s_Ans --> (d,c)] ;
   false.
   
   ?(merge_all_proximals :: (a,b,d,b,c) ==> s_Ans, 0.7, Degree, Result).
   Degree = 0.8,
   Result = [s_Ans --> (a,d,c)] ;
   false.
\end{verbatim}

Due to nontransitivity of proximity relations and the fact that matching finds the first proximal pair (from the left), the order of sequence elements affects the answer. For instance, if we put \verb#a# at the end, we get
\begin{verbatim}
   ?(merge_all_proximals :: (b,d,b,c,a) ==> s_Ans, 0.5, Degree, Result).
   Degree = 0.6,
   Result = [s_Ans --> (d,c,a)] ;
   false.
	
   ?(merge_all_proximals :: (b,d,b,c,a) ==> s_Ans, 0.7, Degree, Result).
   Degree = 0.8,
   Result = [s_Ans --> (d,c,a)] ;
   false.
\end{verbatim}
It happens because $\verb#a#$ and $\verb#c#$ are not close to each (although $\verb#a#$ and $\verb#b#$ as well as $\verb#b#$ and $\verb#c#$ are).
\end{exmp}

\section{Summary}
The main advantages of P$\rho$Log are: compact and declarative
code; capabilities of expression traversal without explicitly programming it; the ability to
use clauses in a flexible order with the help of strategies. Besides, P$\rho$Log has access to the
whole infrastructure of its underline Prolog system. These features make P$\rho$Log suitable for
nondeterministic computations, manipulating XML documents, implementing rule-based
algorithms and their control, etc. \vspace{-0.4cm}
\paragraph{Acknowledgments.} This work has been supported by the Shota Rustaveli National Science Foundation of Georgia under the grant YS-18-1480.

\end{document}